\documentclass[aip,rsi,reprint,graphicx]{revtex4-1} 
\usepackage{graphicx}

\draft 

\begin{document}


\title{Two-layer anti-reflection coating with mullite and  polyimide foam for large-diameter cryogenic infrared filters} 



\author{Yuki Inoue}
\email[]{iyuki@post.kek.jp}
\affiliation{Institute of Physics, Academia Sinica, Nankang, Taipei 11529, Taiwan.}
\affiliation{High Energy Accelerator Research Organization, Ibaraki 305-0801, Japan.}
\author{Takaho Hamada}
\affiliation{Tohoku University Miyagi, Japan.}
\author{Masaya Hasegawa}
\affiliation{High Energy Accelerator Research Organization, Ibaraki 305-0801, Japan.}
\affiliation{SOKENDAI(The Graduate University for Advanced Studies), Ibaraki 305-0801, Japan.}
\author{Masashi Hazumi}
\affiliation{High Energy Accelerator Research Organization, Ibaraki 305-0801, Japan.}
\affiliation{SOKENDAI(The Graduate University for Advanced Studies), Ibaraki 305-0801, Japan.}
\affiliation{Kavli Institute for The Physics and Mathematics of The Universe (WPI), The University of Tokyo, Chiba, Japan.}
\affiliation{Institute of Space and Astronautical Science (ISAS), Japan Aerospace Exploration Agency (JAXA) 
 Kanagawa 252-5210, Japan.}
\author{Yasuto Hori}
\affiliation{University of California, Berkeley, Berkeley CA 94720, USA.}
\author{Aritoki Suzuki}
\affiliation{University of California, Berkeley, Berkeley CA 94720, USA.}
\author{Takayuki Tomaru}
\affiliation{High Energy Accelerator Research Organization, Ibaraki 305-0801, Japan.}
\affiliation{SOKENDAI(The Graduate University for Advanced Studies), Ibaraki 305-0801, Japan.}
\author{Tomotake Matsumura}
\affiliation{Institute of Space and Astronautical Science (ISAS), Japan Aerospace Exploration Agency (JAXA)
 Kanagawa 252-5210, Japan.}
\author{Toshifumi Sakata}
\affiliation{IST corporation, Shiga 520-2153, Japan.}
\author{Tomoyuki Minamoto}
\affiliation{IST corporation, Shiga 520-2153, Japan.}
\author{Tohru Hirai}
\affiliation{Sakane Sangyo Co., Ltd., Osaka 530-0051, Japan}

\date{\today}
\begin{abstract}
We have developed a novel two-layer anti-reflection (AR) coating method 
for large-diameter infrared (IR) filters made of alumina,
for 
the use at cryogenic temperatures in millimeter wave measurements. 
Thermally-sprayed mullite and polyimide foam (Skybond Foam) are used as the AR material. 
An advantage of the Skybond Foam is that the index of refraction is chosen between 1.1 and 1.7 by changing the filling factor.
Combination with mullite is suitable for wide-band millimeter wave measurements with sufficient IR cutoff capability. 
We present the material properties, fabrication of a large-diameter IR filter made of alumina with this AR coating method,
and characterizations at  cryogenic temperatures. 
This technology can be applied to 
a low-temperature receiver system with a large-diameter focal plane 
for next-generation 
cosmic microwave background (CMB) polarization 
measurements, 
such as POLARBEAR-2 (PB-2).
\end{abstract}


\maketitle 

\section{Introduction}
\label{sec:Introduction} 
As astronomical and cosmological observations always demand higher sensitivities, importance of large focal planes with detectors at cryogenic temperatures ever increases. 
The observation of the cosmic microwave background (CMB) polarization, which is one of the best probes for studying the early universe~\cite{Kamionkowski, Seljak}, is an outstanding example where such cryogenic large detector arrays are needed. 
Recently, a few CMB polarization experiments have reported the detection of the faint sub-degree-scale B-mode signal, the odd-parity mode of the CMB polarization pattern~\cite{Seljak}. 
The use of a large array with an order of  10,000 polarization detectors, 
such as the focal plane of 
POLARBEAR-2 (PB-2)~\cite{POLARBEAR2}, SPT-3G~\cite{SPT3G}, and BICEP-3~\cite{BICEP3}, 
is essential for characterizing the B-mode power spectrum. 

The need of a large focal plane leads to a challenge in the thermal design, as large thermal load is expected from a large optical window.~\cite{Hanany:2012vj}
In order to keep the detector sensivitity high enough, efficient infra-red (IR) filters should be considered.
One promising solution to this problem is to 
introduce an alumina plate as 
an IR filter with high thermal conductivity~\cite{Inoue1,Inoue2,POLARBEAR2,SPT3G,BICEP3}. 
The alumina filter absorbs the incident IR emission efficiently. 
The absorbed power is conducted from 
the central region of the filter to the 
edge, which is 
thermally connected 
to 
the cryogenic stage. 
The thermal conductance 
of alumina in the temperature range between 50 and 100~K is 
very high; 
three orders of magnitude 
as large as 
those of conventional plastic filters, such as PTFE, nylon, and black polyethylene.
The temperature gradient and optical loading of the alumina plate are thus very small even with a large diameter, which is ideal as the IR filter material. 

A remaining technological challenge on the use of alumina IR filter is to establish AR coatings for a large frequency coverage. 
Alumina is known to be highly reflective in the millimeter wavelength. 
We thus need an 
AR coating on each surface of the alumina disc.
AR coating on such a large surface
is easily peeled off owing to the thermal shock at cold 
temperatures. 
The aforementioned next-generation CMB projects demand multi-layer AR coatings on large IR filters with a typical diameter of 500~mm. 
In the previous studies, the POLARBEAR group 
attempted to develop 
a two-layer AR coating method with epoxy on an alumina surface, with a diameter of 50~mm~\cite{Inoue1,Inoue2,Rosen:2013zza}.
However, the epoxy and alumina cracked during thermal cycling.
To avoid the cracking, 
stress-relief grooves were adopted to reduce the stress between alumina and the epoxy layer.~\cite{Inoue1} 
This technology was 
also used in 
the BICEP-3~\cite{BICEP3} group, 
which succeeded in making an single-layer AR coating with epoxy for a diameter of about $500~\mathrm{mm}$.
However, this method is expensive. Also, it is not easy to maintain the thickness uniformity of the AR layers on the thin alumina plate. 
Alternatively, 
the ACTpol~\cite{ACTpolAR} group succeeded in making two-  and three-layer AR coatings on silicon lenses, whose diameters were approximately 300 mm.
They tried and succeeded using the technology of 
subwavelength grating (SWG). 
It is theoretically possible to apply this technology to an alumina surface; however, it is difficult because the dicing blade is subject to wear 
so that 
groove pitch and depth are changed.
It is 
also 
difficult to make a large ingot of 
silicon. Machining the material is also very expensive. 
Yet another method studied for the SPIDER project is 
a polyimide sheet as an AR coating on 
the 
sapphire 
with a diameter of about 250~mm \cite{Sean}. 
The SPIDER method is very easy to perform and less expensive than 
the method 
with epoxy or SWG. 

Finally, the first application of thermal sprayed mullite as AR coating
was made on a 50 mm diameter alumina disc~ \cite{Toki:2013}.
It is also possible to tune the dielectric constant of a thermal sprayed layer
by spraying microspheres and alumina powder with different ratios.
A three-layer thermal sprayed broadband AR coating that achieves
over a 100 \% fractional bandwidth has been demonstrated on a 50 mm diameter alumina disc~\cite{Oliver}..

In this study, we newly employ the polyimide foam coating on the mullite coating described above to establish a two-layer AR coating method. We apply this technique to a thin and large alumina disc with a 460 mm diameter as an IR filter, and demonstrate the performance of the AR coating.




We fabricated a 2 mm thick and 460 mm diameter alumina filter, and demonstrated the performance of the new AR coating.
Our development focuses on the use in future CMB measurements but the method developed can also be used for other applications. 
In Section~\ref{sec:FilterDesign} we describe the design of a large-diameter alumina IR filter with our new AR coating method. Section~\ref{sec:MaterialProperties} explains our measurements of material properties. The fabrication process is detailed in Section~\ref{sec:Fabrication}. Section~\ref{sec:Characterization} shows our characterization of the AR coatings on a large-diameter alumina IR filter. We discuss our results in Section~\ref{Disc} and give conclusions in Section~\ref{sec:Conclusion}. 

\section{Design} 
\label{sec:FilterDesign} 
To be concrete, our AR-coating deveopment targets next-generation CMB experiments, where
typical requirements on the IR filter and the AR coating are as follows: \cite{footnote1} 
\begin{itemize}
	\item The filter diameter must be extendable to 450~mm or larger;
	\item The AR coating is robust against thermal shock, so that it is usable in cryogenic temperatures; 
	\item The uniformity of the additional optical path length due to the AR coating should 
	be less than $\lambda/50$, where $\lambda$ is the wavelength of the incoming electromagnetic wave;
	\item Transmittance should be above 95~\% at the detection bands, which we assume 95~GHz and 150~GHz with a $30~\%$ fractional bandwidth for each; 
	\item The 3~dB cutoff frequency should be below 1~THz. 
\end{itemize}
The filter diameter requirement satisfies the requirement for a large focal plane with O(10000) superconducting millimeter-wave sensors. 
The requirement on robustness against thermal shock is important for use at low temperatures. 
The thickness of the AR coating is designed to be $\lambda/4$ so that the reflected beam is 
canceled. 
The uniformity must be 
better 
than $\lambda/50$, which corresponds to a transmittance uncertainty of 1~\%.  
The requirement on transmittance in detection bands comes from the design sensitivity of experiment.   
The requirement on the cutoff frequency is derived to reduce the IR emission whose main frequency range is 10-100~THz. 

\begin{figure}
\includegraphics[width=6.6cm]{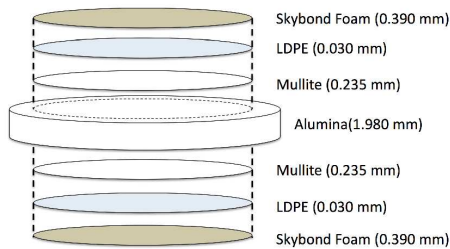}
\caption{A schematic view of the 
two-layer AR coatings on a large-diameter alumina disc. 
The AR coatings consist of the thermally-sprayed mullite as the first layer and polyimide foam (Skybond Foam) as the second layer. 
A thin layer of low-density polyethylene (LDPE) is placed in between two AR layers as a glue layer. 
\label{fig:af_config}}
\end{figure}

Figure~\ref{fig:af_config} shows a schematic view of two-layer AR coatings on a large-diameter alumina disc. The design satisfies the typical requirements mentioned above. The AR coatings consist of the thermally-sprayed mullite, which has the index of refraction (IOR) of 2.5 as the first layer and polyimide foam (called Skybond Foam~\cite{IST} hereafter) as the second layer. A thin layer of low-density polyethylene (LDPE) is placed in between two AR layers as a glue layer. 
We employ the mullite as the first layer because the coefficient of thermal expansion (CTE) is close to that of alumina. 
The idea of using the Skybond Foam for the second layer is a natural extension of the polyimide sheet adopted by SPIDER, which was explained in the previous section. 
An important advantage of the polyimide foam is that the IOR can be controlled by changing the filling factor of the polyimide, which gives us an additional tuning knob to optimize the AR coating properties for given observation frequencies. 
Our example design for this study is optimized to achieve the highest integrated efficiency for simultaneous measurements centered at 95~GHz and 150~GHz with a $30~\%$ fractional bandwidth for each. Here we fix the thickness of the alumina and the IOR of mullite, while the IOR of the Skybond Foam and the thicknesses of mullite and the Skybond Foam layers are free parameters. For the design in Fig.~\ref{fig:af_config}, the optimized IOR for the Skybond Foam is 1.43.
Measurements of the properties of each AR layer as well as those of alumina we have used in this study are described in the next section. 

\section{Material properties} 
\label{sec:MaterialProperties} 

We have measured basic properties of alumina, mullite and the Skybond Foam. The measurements include thicknesses, IORs, and the loss tangent values. Table~\ref{material} summarizes our results. 

The IOR and the loss tangent are obtained from transmission measurements. 
A detailed description of the measurement system is available elsewhere~\cite{Inoue1}. 
The beam at millimeter wavelength (72-108~GHz and 108-148~GHz) is produced by a signal generator with a sixfold and eightfold frequency multiplier\cite{AMT}. 
The frequency range of the signal generator is between 12 and 18~GHz.
The beam is collimated by an ultra-high-molecular-weight polyethylene lens and the measured sample is placed immediately after the aperture.
The transmitted radiation is detected by a diode detector. 
The signal is chopped at 13 Hz for modulation with a lock-in amplifier, and the modulated signal is obtained using a DAQ. 
The detector is moved along the optical axis for more than a half wavelength to subtract the effect of a standing wave in the measurement setup. 
We measure the power with and without the sample and take the ratio between them to obtain the transmittance.
The measured transmittance is fitted with the Fresnel formula~\cite{Hecht}.
In the following sections, we describe more details on each material. 

 \begin{table} 
  \caption{ Basic properties of alumina and AR materials. 
The thickness $d$, the IOR $n$, and the loss tangent $\tan\delta$  are listed. 
  Our measurements for $n$ and $\tan\delta$ are 
  carried out 
  both at room temperature and 
  at the temperature of liquid nitrogen,
whereas the thickness values are 
only measured 
at room temperature. 
The errors include both statistical and systematic uncertainties. The systematic errors 
mainly 
arise from the uncertainty of the AR thicknesses and 
temperature variations during the measurements. 
\label{material}} 
 \begin{tabular}{c|c|c|c|c} \hline
    Material &T& $d$ & $n$ & $\tan\delta$ \\ 
    &[K]&[mm]&& [$\times 10^{-4}$] \\ \hline
    Alumina &298& $1.98\pm0.01$ & $ 3.144\pm0.005$ & $3.7\pm1.7$   \\
    Mullite &298& $0.24\pm0.02$ & $ 2.52 \pm 0.02$  & $121\pm16$ \\
    Skybond Foam &298& $0.39\pm0.02$ & $ 1.436\pm 0.025$  &  $120\pm2$ \\ \hline
       Alumina &81& $-$ & $ 3.117\pm0.005$ & $3.0\pm1.1$   \\ 
    Mullite &81& $-$ & $ 2.46\pm 0.03$  & $53\pm10$ \\ 
    Skybond Foam &81& $-$ & $ 1.425\pm 0.025$  &  $25\pm1$ \\ \hline 
  \end{tabular}
\end{table}

\subsection{Alumina} 
The sintered polycrystalline alumina was manufactured by Nihon Ceratech corporation with a purity of 99.5~\% (labeled as 99.5LD)~\cite{Nihon}. 
The loss tangent of the 99.5~\% purity of the alumina is low in the detection band and high in the sub-millimeter wavelength.
The large gradient of the loss tangent between the detection band and the sub-millimeter region results in
a sharp IR cutoff. We have also tested alumina with purities of 99.9~\% and 99.99~\% from the same company and
have concluded that 99.5LD is the best material.
More details are described elsewhere \cite{Inoue2}.

\subsection{Mullite} 
\begin{figure}
\includegraphics[width=9cm]{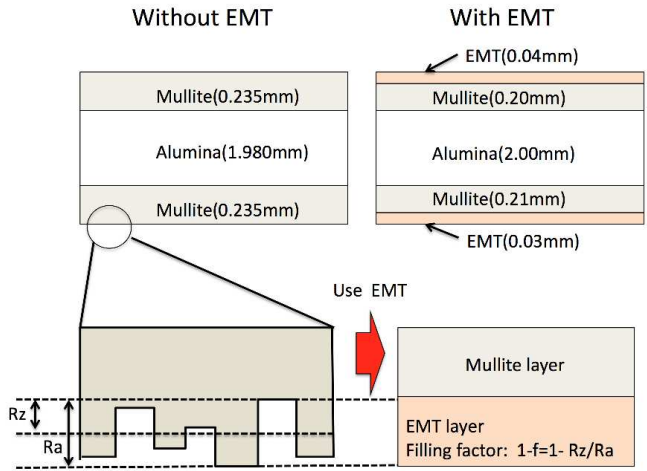}
\caption{The schematic view of 
the mullite surface and the model with the effective medium theory (EMT). 
The mullite has an asperity 
surface, which we can approximate 
as the virtual layer based on the EMT. The effective IOR of the virtual layer is described by Equation~(\ref{eq:EMT}). This equation depends on the filling factor which is given by the ratio between $R_{\mathrm{a}}$ and $R_{\mathrm{z}}$.   \label{fig:model_mullite}}
\end{figure}

Mullite is a high-performance ceramic stable at atmospheric pressure. We place the mullite on both alumina surfaces using the thermal-spraying method developed by Tocalo corporation~\cite{tocalo,Toki:2013} 
We measure the transmittance of the mullite at 298~K and 81~K.
The measured frequencies are between 72 and 148~GHz. 
This measurement yields fringe patterns 
from the interferogram of the alumina and mullite layers.
Mullite has asperity structure, with a surface roughness $R_a$ and a ten-point average roughness $R_z$ of $6~\mathrm{\mu m}$ ($6~\mathrm{\mu m}$) and $36~\mathrm{\mu m}$ ($34~\mathrm{\mu m}$) at one (the other) side, respectively.
We apply the effective medium theory (EMT)~\cite{EMT} to the surface of mullite 
to obtain the effective IORs as follows; 
\begin{equation}
n_{\mathrm{eff}}=n_\mathrm{m} \sqrt{\frac{2(1-f)(n_0^2-n_{\mathrm{m}}^2)+n_0^2+2n_{\mathrm{m}}^2}{2n_{\mathrm{m}}^2+n_0^2 + (1-f)(n_{\mathrm{m}}^2-n_0^2)}}, \label{eq:EMT}
\end{equation}
where $n_{\mathrm{eff}}$, $n_{\mathrm{m}}$, and $n_0$ are the effective IOR, IOR of mullite and vacuum, respectively, and $(1-f)$ is the filling factor, defined as $1-R_{\mathrm{a}}/R_{\mathrm{z}}$.
We can approximate the asperity surface as the virtual layer from the EMT as 
illustrated 
in Fig.~\ref{fig:model_mullite}.
The estimated IORs and loss tangents 
in Table.~\ref{material} are obtained using the EMT. 



\subsection{Skybond Foam} 
\begin{figure}
\includegraphics[width=9cm]{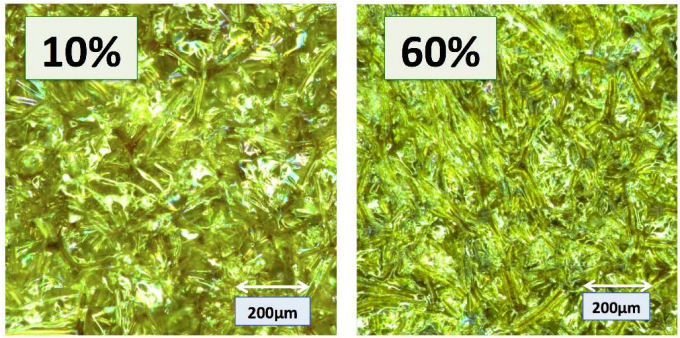}
\caption{Left: Photomicrograph of Skybond Foam with a 10~\% filling factor. 
The bubbles are pressed with a 10 ton weight. 
The average diameter of the unit cells is 
$10~\mathrm{\mu m}$.   
Right: Measured sample of Skybond Foam with a 60~\% filling factor. \label{fig:micro}}
\end{figure}

\begin{figure}
\includegraphics[width=9cm]{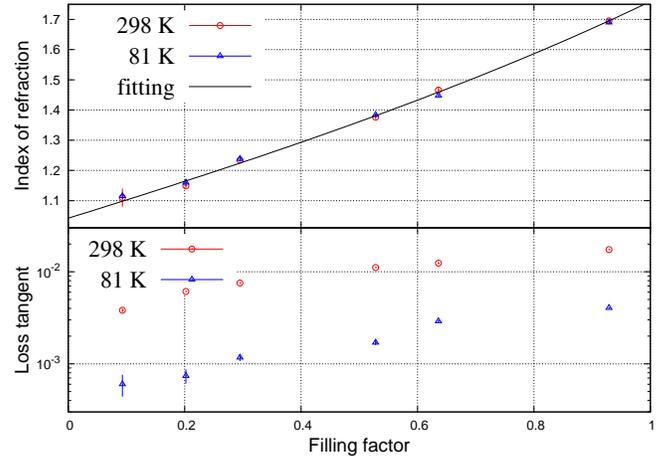} 
\vspace{1cm} 
\caption{The IORs (top) and loss tangents (bottom) as a function of filling factor. The sample is placed at 298~K or 81~K. The error bars in the $y$ axis correspond to systematic error of millimeter source fluctuation. The size of the measured samples are $100~\mathrm{mm} \times 100~\mathrm{mm} \times 10~ \mathrm{mm} $. \label{fig:EMT}}
\end{figure}

The Skybond Foam is an polyimide foam, which is manufactured by the IST corporation~\cite{IST}.
Figure~\ref{fig:micro} shows photomicrographs of the 
Skybond Foam with a filling factor of 10~\% and 60~\%. 
The filling factor is 
defined as 
the volume-filling fraction of the polyimide  in the mixture, which can be controlled from 10~\% to 100~\%. 
Each cell of the Skybond Foam has a size of $\sim10~{\mathrm{\mu m}}$.
In general, the IORs and loss of the material depend on the filling factor~\cite{filling}. 
We measure the IOR and loss tangent of Skybond Foam with 
different filling factors 
between 10 and 100~\% (10,  20, 30, 50, 60, and 100~\%).
The size of each sample is $100~\mathrm{mm} \times 100~\mathrm{mm}$ with a thickness of 10~mm. 
The thick Skybond Foam was chosen to perform measurements with high signal-to-noise ratios. 
The measured IOR and loss tangent at 298K and 81 K are shown in Fig.~\ref{fig:EMT}.
The black curve is the best 
fit for the 81 K measurement, where the EMT with Eq.~(\ref{eq:EMT}) is used. 
The estimated 
IORs are $n_0=1.049 \pm 0.005 $ and $n_m=1.748 \pm 0.006 $. 
We decide to use 
a filling factor of 60 \% 
for the AR material because 
it gives the IOR closest to the target value of 1.43. 


\begin{figure}
\includegraphics[width=9cm]{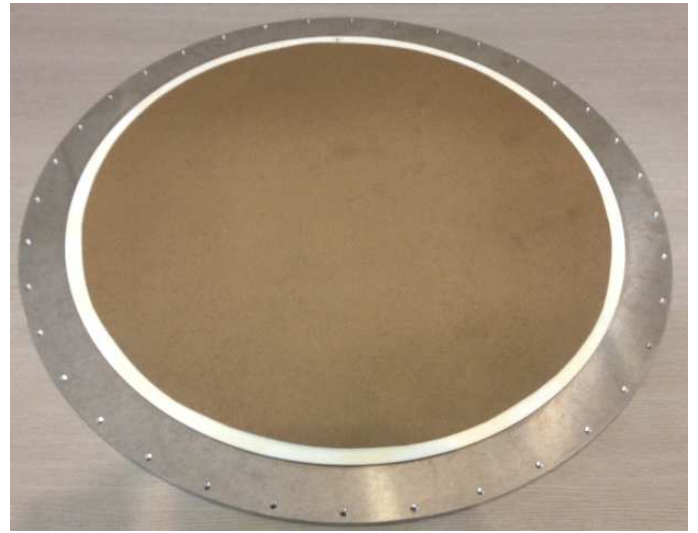}
\caption{\label{fig:alumina_filter} An alumina filter with a two-layer AR coating 
on 
both surfaces. 
The diameter of the alumina disc 
is 460~mm.
The diameter of the AR layers is 440~mm. 
The 10~mm edge part of the alumina discs is without the AR coatings and is used to have a direct thermal contact.} 
\end{figure}

\section{Fabrication}
\label{sec:Fabrication} 
Figure~\ref{fig:alumina_filter} shows a photo of the fabricated alumina IR filter with the AR coating with mullite and Skybond Foam. 
We use an alumina disc with a diameter of 460~mm and a thickness of 2~mm. 
We place the mullite layer with thermal spraying on both sides of the alumina. The thickness of the mullite is 
$(0.24\pm0.02)$~mm 
as 
already listed 
in Table~\ref{material}.
We place 
a sheet of low-density polyethylene (LDPE) with a thickness of $30~\mathrm{\mu m}$ as the bond layer. 
This layer 
is 
sandwiched 
by 
the Skybond Foam and mullite surface. 
The thickness of the Skybond Foam is $(0.39 \pm 0.02)~\mathrm{mm}$
as already listed in Table~\ref{material}. 
The surface of Skybond Foam is then 
shaped with 
a 
milling machine, whose diameter and rotating speed of 
the 
blade are 4~mm and 6000~rpm, respectively.
The uniformity of the thickness is improved by a factor of two through the machining process. 
The stack of Skybond Foam, mullite and alumina layers are 
then 
baked in an oven.
We sandwich the stacked layer with 
10~mm thick aluminum 
boards 
and press 
them 
using lead blocks with a pressure of $3,000~\mathrm{N/m^2}$.  
We adopt the baking recipe established by the SPIDER group~\cite{Sean}. 
We start the 
baking 
cycle by slowly ramping up 
the temperature 
from room temperature to 
the baking temperature of
160 $^\circ$C 
over 6 hours to avoid thermal shock.
After keeping the baking temperature for 6 hours, 
we spend another 6 hours to ramp down to room temperature. 

\section{Characterization} 
\label{sec:Characterization} 

We perform a series of measurements to verify that the fabricated large-diameter sample satisfies each requirement listed in Section~\ref{sec:FilterDesign}. 
First of all, we run a thermal cycle test from 300~K to 77~K with conductive cooling.
The AR coating passes the thermal cycle test over ten times. 
No damage such as cracks due to thermal contraction is found.

\begin{figure}
\includegraphics[width=9cm]{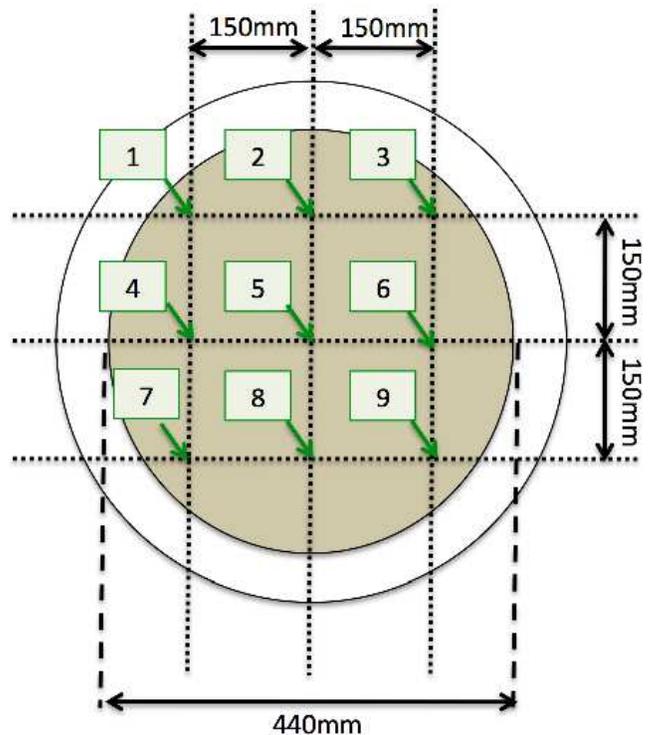}
\caption{Definition of the points of uniformity measurements. The point number 5 is at the center of the sample. We measure transmittances at the 9 points for estimation of the IOR. The measured IOR's are shown in Fig.~\ref{fig:uni}.  \label{fig:uni_def}}
\end{figure}

\begin{figure}
\includegraphics[width=9cm]{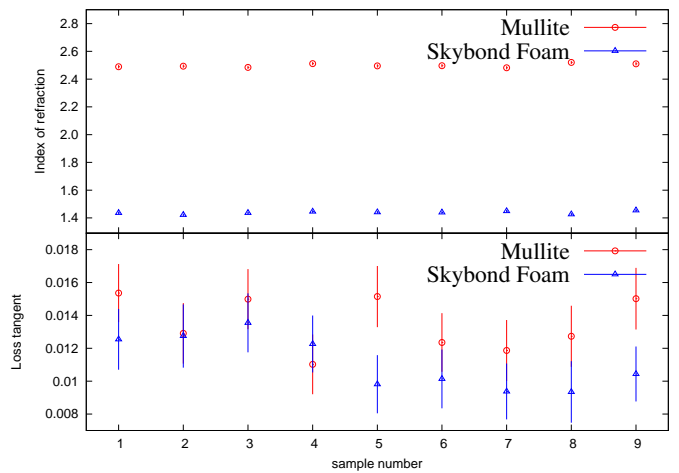} 
\vspace{1cm} 
\caption{Uniformity of IORs (top) and loss tangents (bottom) 
for the 9 measurement points. 
The Skybond Foam and mullite are measured with 
the millimeter wave source with a frequency range from 72 to 108~GHz. The sample is placed at 298~K. The error bars correspond to statistical uncertainties.  \label{fig:uni}}
\end{figure}

\begin{figure}
\includegraphics[width=9cm]{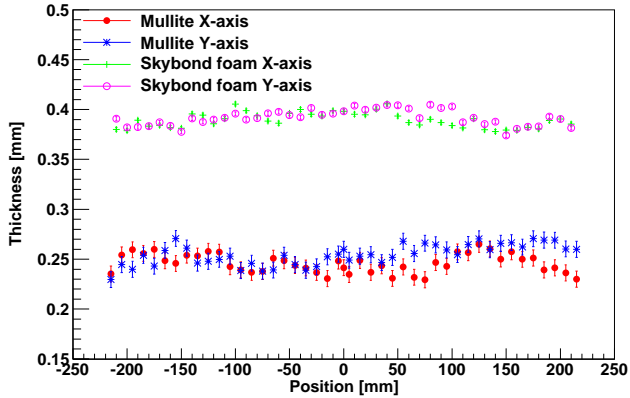}
\caption{Thicknesses of the Skybond Foam and mullite layers which are characterized by CMM to a precision of $2~\mathrm{\mu m}$. 
The red and blue points are the characteristic directions of the X and Y-axes, respectively. The each position corresponds to the distance from the center of the sample. The acceptable thickness uncertainties of Skybond Foam and mullite are 0.037~mm and 0.022~mm, respectively. The error bars correspond to systematic errors of the thickness rms of alumina. These results meet our requirements.  \label{fig:CMM}}
\end{figure}

To check the uniformity, we select 9 positions shown in Fig.~\ref{fig:uni_def} and measure the IOR and loss tangent at room temperature for a frequency range between 72 and 108~GHz. The instrument described in Section~\ref{sec:MaterialProperties} is used for the measurements. The results are shown in Fig.~\ref{fig:uni}. 
We 
then 
measure 
the thickness uniformity of the Skybond Foam and mullite using a coordinate measuring machine (CMM). The results are shown in Fig.~\ref{fig:CMM}. We calculate the optical path length in both Skybond Foam and mullite with the measured IOR and thickness. These fluctuations are less than $\lambda/50$ as shown in Table~\ref{opl}. 


 \begin{table} 
  \caption{Measured uniformities of the Skybond Foam and mullite. 
  The mean and rms 
  of the thickness, $d$, and IOR, $n$, 
  are calculated from the measurements 
  shown in Figs.~\ref{fig:uni} and \ref{fig:CMM}.  \label{opl}} 
 \begin{tabular}{c|c|c|c|c|c} \hline
     &\multicolumn{2}{c|}{$d$ [mm]}  & \multicolumn{2}{c|}{$n$}  &Optical path \\ 
    &mean&rms&mean&rms&length [mm] \\ \hline
    Mullite &0.250& 0.010 &  2.46 & 0.063& 0.016\\
    Skybond Foam &0.391& 0.008 & 1.43  &  0.020&0.009 \\ \hline
  \end{tabular}
\end{table}

\begin{figure}
\includegraphics[width=8.5cm]{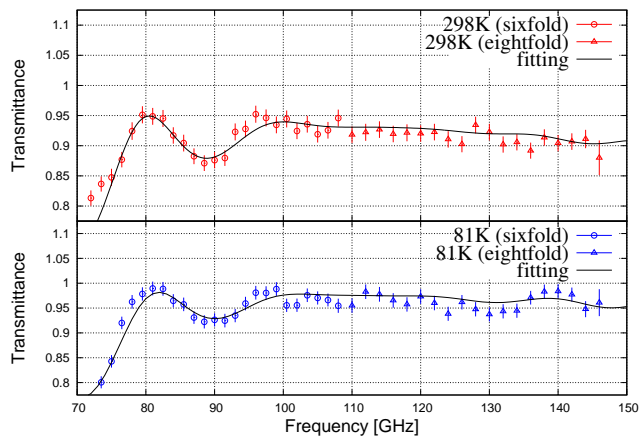}
\vspace{1cm} 
\caption{The transmittances of the IR filter 
the function 
of the frequency at 298~K (top) and 81~K (bottom). Two frequency multipliers (sixfold and eightfold) are used to cover the entire frequency range. \label{fig:AR}}
\end{figure}

We measure the transmittance of the alumina filter with the same setup described in 
Section~\ref{sec:MaterialProperties} 
at room temperature and at 81~K cooled with liquid nitrogen.
Since the uniformity measurements at room temperature gurantee that the
fabrication process is independent of the size of the filter, to perform measurements with liquid nitrogen, we fabricated a sample with 50~mm in diameter to avoid mechanical conflict with the measurement system. 
Figure~\ref{fig:AR} shows the result.
The black lines show fit results with the EMT. 
The estimated 
average 
transmittances are 95.9~\% and 95.8~\% in the 95~GHz and 150~GHz bands, respectively. 
We estimate the emissivity, $\epsilon$, from the following equation:  
\begin{equation}
	\epsilon= A =1-R( d, n, \tan{\delta}, \nu)-T( d, n, \tan{\delta}, \nu),
\end{equation}
where $A$, $R$ and $T$ are the absorption, the power reflectance, and the transmittance of the filter, respectively. 
The reflectances calculated from the parameters given in Table~\ref{material} are 2.1~\% and 1.1~\% at 95~GHz and 150~GHz bands. 
The estimated emissivities are 2.7~\% and 4.2~\% in the 95~GHz and 150~GHz bands, respectively.

The transmittance of the alumina filter at sub-millimeter wavelengths is measured using a Fourier transform spectrometer (FTS) with 
an 
InSb bolometer.
A mercury lamp is used as a 
broadband thermal source. 
The detailed description of 
the 
FTS is found elsewhere~\cite{1998PASJ...50..359M}.
 The transmittance is measured from 200~GHz to 1,500~GHz at two different temperatures of 19 K and 298 K.
 We prepare the alumina filters with a diameter of 20~mm for the FTS measurements. 
 
\begin{figure}
\includegraphics[width=9.5cm]{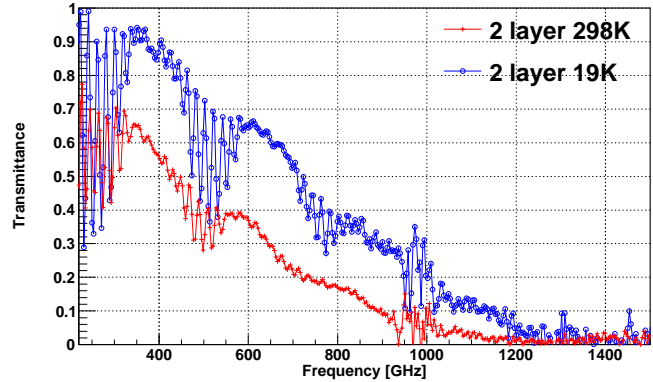} 
\caption{The transmittance of the IR filter at the THz band. The red crosses represent measurements at 298~K and the blue open circles are for 19~K.\label{fig:IR}} 
\end{figure}

The measured spectrum at 19~K has a higher transmittance than the spectrum at 
298~K as shown in Fig.~\ref{fig:IR}. 
The 3~dB cutoff frequencies of the filter are estimated as 400 and 650~GHz at 295 and 19~K, respectively. 
These 3~dB cut off frequencies are sufficiently lower than our design goal. 

\section{Discussion \label{Disc}} 


We have demonstrated that the AR coating method with mullite and Skybond Foam satisfies all the requirements listed in Section~\ref{sec:FilterDesign}, which are typical in next-generation CMB experiments. 
Although we have focused on the measurements at 95 GHz and 150 GHz, a future CMB satellite mission LiteBIRD~\cite{Hazumi:2012aa,Matsumura:2013aja,Matsumura2016} for example requires wider frequency ranges from 40 GHz to 400 GHz using two telescopes. 
Skybond Foam is tunable in IOR and can thus be considered as a candidate technology for such a space mission. 
We have newly applied the EMT in millimeter wave measurements to have the better surface model, which were not adopted in previous works.
The fit results are in better agreement with data. 
The performances of the new AR coating method described in this study are as good as those with epoxy with grooves.
The fabrication time of the method with mullite and Skybond Foam is much shorter than the epoxy coating, which is a clear advantage.
Finally, this method can in principle be applied to curved surfaces, so that applications for lenses should also be considered.

\section{Conclusion}
\label{sec:Conclusion} 
We have newly introduced polyimide foam (Skybond Foam) as the AR coating material. 
We show that the IOR of Skybond Foam is controlled between 1.1 and 1.7 by changing the filling factor. The tunable IOR allows us to design a multi-layer AR coating much easier. 
We have made a two-layer AR coating with thermally-sprayed mullite as the first layer and Skybond Foam as the second layer on a 2 mm-thick alumina plate with 460 mm in diameter. We have confirmed that sample uniformity in the IOR and thickness is satisfactory for 150 GHz measurements. We have also confirmed that the AR coating is robust against thermal cycles. 
The measured transmittances of this filter 
at 81 K are 95.8~\% for 95 GHz and 95.9~\% for 150 GHz. 
The measured 3~dB cut-off 
frequency is 650 GHz at 19 K. 
The results satisfy requirements on IR filters for next-generation CMB polarization measurements where a large focal-plane and a large window are needed. 

\begin{acknowledgments}
We thank Hiroshi Matsuo and Tom Nitta for their help 
with THz measurements.
We are also 
grateful 
to Satoru Igarashi, Iwao Murakami, and Keishi Toyoda for their help in setting up the measurement.
We would like to express our 
gratitude to Adrian Lee, Osamu Tajima, Suguru Takada,  Akito Kusaka, Hiroshi Yamaguchi, Junichi Suzuki, Yuichi Watanabe and Yuuko Segawa.
We would like to thank the KEK Cryogenics Science Center for the support. 
Yuki Inoue 
was supported by Advanced Research Cource in SOKENDAI (The Graduate University for Advanced Studies), 
and 
by Academia Sinica under Grants No. CDA-105-M06 in Taiwan.
KEK authors were supported by MEXT KAKENHI Grant Numbers 21111002, 15H05891, and JSPS KAKENHI Grant Numbers 13J03626, 24740182, 24684017,   15H03670, 24111715 and 26220709.
Finally, this work was supported by the JSPS Core-to-Core Program, A. Advanced Research Networks.

\end{acknowledgments}


\begin{thebibliography}{99}
\bibitem{Kamionkowski}  M.Kamionkowski, A.Kosowsky and A.Stebbins, ``A Probe of Primordial Gravity Waves and Vorticity,'' Phys. Rev. Lett. $\bm{78}$, 2058-2061(1997).
\bibitem{Seljak}U. Seljak and M. Zaldarriaga,``Signature of Gravity Waves in the Polarization of the Microwave Background,'' Springer US. $\bm{78}$, 2054 (1997).

\bibitem{POLARBEAR2}A. Suzuki {\it et al.}, ``The POLARBEAR-2 Experiment,'' Journal of Low Temperature Physics $\bm{176}$, 719-725~(2014).

\bibitem{SPT3G} B. A. Benson {\it et al.} ,``SPT-3G: A Next-Generation Cosmic Microwave Background Polarization Experiment on the South Pole Telescope,'' Proc.SPIE $\bm{9153}$, 91531P~(2014).


\bibitem{BICEP3} Z. Ahmed {\it et al.}, ``BICEP3: a 95 GHz refracting telescope for degree-scale CMB polarization,'' Proc.SPIE $\bm{9153}$, 91531N~(2014).

\bibitem{Hanany:2012vj}  S.~Hanany, M.~Niemack and L.~Page,``CMB Telescopes and Optical Systems,"   Springer Science+Business Media Dordrecht ,431(2013).  

\bibitem{Inoue1} Y.~Inoue {\it et al.},'' Appl. Opt. $\bm{53}$, 1727-1733, (2014).

\bibitem{Inoue2} Y.~Inoue {\it et al.}, "Thermal and optical characterization for POLARBEAR-2 optical system", Proc.SPIE $\bm{9153}$, 91533A~(2014).

\bibitem{Rosen:2013zza} D. Rosen {\it et al.}, ``Epoxy-based broadband anti-reflection coating for millimeter-wave optics, Appl. Opt. $\bm{52}$, 8102-8105, (2013).

\bibitem{ACTpolAR} R.~Datta {\it et al.},``Large-aperture wide-bandwidth antireflection-coated silicon lenses for millimeter wavelengths,'' Appl. Opt. $\bm{52}$, 8747-8758 (2013).

\bibitem{Sean} B. Sean, ``Half-wave Plates for the Spider Cosmic Microwave Background Polarimeter," Ph.D. thesis, Department of Physics CASE WESTERN RESERVE UNIVERSITY, (2014).

\bibitem{Toki:2013} A. Suzuki, ``Multichroic Bolometric Detector Architecture for Cosmic Microwave Background Polarimetry Experiments," Ph.D. thesis, University of California, Berkeley, (2013).

\bibitem{Oliver}O. Jeong {\it et al.},``Broadband Plasma-Sprayed Anti-reflection Coating for Millimeter-Wave Astrophysics Experiments,'' Journal of Low Temperature Physics, pp 1-6(2016).

\bibitem{footnote1} These mainly come from requirements from the POLARBEAR-2 experiment.

\bibitem{IST} I.S.T Corporation, \url{http://www.istcorp.jp/}.

\bibitem{AMT} The part number and output frequency of synthesized frequency generator are Agillent 83711B and 12-18 GHz, respectively. The frequency multiplier is made by AmTechs Corporation, \url{http://www.amtechs.co.jp}.
\bibitem{Hecht}E. Hecht, Optics fourth edition (Addison Wesley, 2002).

\bibitem{Nihon} Nihon Ceratech, \url{http://www.ceratech.co.jp}.

\bibitem{tocalo} Tocalo, \url{http://www.tocalo.co.jp}.

\bibitem{EMT}C. Tuck, Effrctive Medium Theory (Oxford: Clarendon Press 1999).

\bibitem{filling} O. Levy. and D.Stroud. ``Maxwell Garnett theory for mixtures of anisotropic inclusions: Application to conducting polymers." Physical Review B, $\bm{56(13)}$, 8035~(1997).
\bibitem{1998PASJ...50..359M} H. Matsuo, A. Sakamoto, and S. Matsushita, ``FTS Measurements of Submillimeter-Wave Atmospheric Opacity at Pampa la Bola", PASJ $\bm{50}$, 359-366~(1998).



\bibitem{Hazumi:2012aa} M. Hazumi et al., Proc. SPIE Int. Soc. Opt. Eng. 8442-844219 (2012).

\bibitem{Matsumura:2013aja} 
  T.~Matsumura {\it et al.},
  J.\ Low.\ Temp.\ Phys.\  {\bf 176}, 733 (2014)
  doi:10.1007/s10909-013-0996-1
  [arXiv:1311.2847 [astro-ph.IM]].
  
\bibitem{Matsumura2016}
  T.~Matsumura {\it et al.}, 
  Journal of Low Temperature Physics (2016), 10.1007/s10909-016-1542-8.

\end{thebibliography}
\end{document}